\documentclass[onecolumn,traditabstract]{aa}

\usepackage{natbib}
\usepackage{amssymb,amsfonts}
\usepackage{amsmath,mathtools}
\usepackage{url}
\usepackage{hyperref}

\begin{document}

\title{Near optimal angular quadratures for polarised radiative transfer\footnote{The tables mentioned in Appendix \ref{sec:app} can be found in electronic form at the CDS via anonymous ftp to \url{cdsarc.u-strasbg.fr} (130.79.128.5)
or via \url{http://cdsweb.u-strasbg.fr/cgi-bin/qcat?J/A+A/}}}

\author{Ji\v{r}\'{\i} \v{S}t\v{e}p\'an\inst{1} \and Jaume Jaume Bestard\inst{2,3} \and Javier Trujillo Bueno\inst{2,3,4}}

\institute{
Astronomical Institute ASCR, v.v.i., Ond\v{r}ejov, Czech Republic.
\email{jiri.stepan@asu.cas.cz}
\and
Instituto de Astrof\'{\i}sica de Canarias, V\'{\i}a L\'actea s/n, E-38205 La Laguna, Tenerife, Spain. \email{jjaumeb@iac.es}
\and
Departamento de Astrof\'{\i}sica, Universidad de La Laguna (ULL), E-38206 La Laguna, Tenerife, Spain
\and
Consejo Superior de Investigaciones Cient\'{\i}ficas, Spain. \email{jtb@iac.es}
}

\date{Received XXXX / Accepted XXXX}

\abstract{
In three-dimensional (3D) radiative transfer (RT) problems, the tensor product quadratures are generally not optimal in terms of the number of discrete ray directions needed for a given accuracy of the angular integration of the radiation field. In this paper, we derive a new set of angular quadrature rules that are more suitable for solving 3D RT problems with the short- and long-characteristics formal solvers. These quadratures are more suitable than the currently used ones for the numerical calculation of the radiation field tensors that are relevant in the problem of the generation and transfer of polarised radiation without assuming local thermodynamical equilibrium (non-LTE). We show that our new quadratures can save up to about 30\,\% of computing time with respect to the Gaussian-trapezoidal product quadratures with the same accuracy.
}


\titlerunning{Angular quadrature for polarised radiative transfer}
\authorrunning{J. \v{S}t\v{e}p\'an et al.}

\maketitle


\section{Introduction\label{sec:intro}}

The non-linear radiative transfer problem out of local thermodynamic equilibrium (hereafter, non-LTE) must be solved by an iterative solution of the equations of statistical equilibrium (ESE) and the radiative transfer equation (RTE). In most applications in astrophysics, the numerical solution of the RTE is the most computationally demanding part of the problem. This is especially true for the non-axisymmetric and generally multi-dimensional problems.

In the radiative transfer codes based on the concept of short or long characteristics on Cartesian grids, the discrete set of rays along which the formal solution of the RTE is calculated are the angular quadrature points on a unit sphere. The values of Stokes parameters at the quadrature points are used, with appropriate weights, to obtain the scalar mean intensity $J$ or the radiation field tensors $J^K_Q$ in the general polarisation transfer case \citep[see e.g.][]{ll04}. Since these quantities enter the ESE, their accuracy is crucial for the accuracy of the atomic density matrix elements and, consequently, for the accuracy of the whole non-LTE solution.

In RT problems involving three-dimensional (3D) model atmospheres and scattering polarisation, little is known about the suitability of different quadrature rules. In general, the higher the number of rays is in the quadrature, the better the accuracy of the $J^K_Q$ tensor integration is. On the other hand, the overall computing time (CPU time) increases approximately proportionally to the number of ray directions. This becomes especially important for solving large-scale problems in 3D models resulting from realistic radiation-magnetohydrodynamical simulations where the CPU time for one formal solution can be of the order of tens of thousands of hours \citep[e.g.][]{stepan16}.

Our goal in this paper is to find an angular quadrature for the solution of the transfer of polarised radiation that guarantees any given accuracy with the minimum number of points, that is, an optimal quadrature. While recently, significant progress has been made in the mathematical research on optimal quadratures in general, information was still lacking as to the specific field of polarised radiative transfer. This paper tries to fill in this gap. Since the quadratures we derived are optimal or near optimal for the problems arising in the theory of polarised radiative transfer, they are generally more efficient than the currently used quadratures. In Sect.~\ref{sec:formul} we formulate quantitatively the problem of optimal quadrature for spectropolarimetry. In Sect.~\ref{sec:calc} we derive the key system of equations and we discuss their numerical solution. In Sect.~\ref{sec:results} we present our results for various levels of accuracy. Finally, in Sect.~\ref{sec:concl} we summarise our findings.


\section{Formulation of the problem\label{sec:formul}}

The integral of a real or complex-valued function $f(\vec x)$ over the unit sphere $\mathbb{S}^2=\{\vec\Omega\in\mathbb{R}^3: \|\vec\Omega\|=1\}$ can be approximated by a weighted sum of the function values sampled at a finite number of $N$ discrete points $\{\vec\Omega_i\}_{i=1}^N$ on $\mathbb{S}^2$. In case the quadrature is exact for the function $f(\vec\Omega)$,
\begin{equation}
\int_{\mathbb{S}^2} f(\vec\Omega)\; d\vec\Omega=\sum_{i=1}^N w_i f(\vec\Omega_i)\,,
\label{eq:qdef}
\end{equation}
where $\{w_i\}_{i=1}^N$ are positive weights associated with the corresponding unit vectors $\vec\Omega_i$. The quadrature $\{w_i,\vec\Omega_i\}_{i=1}^N$ is said to be optimal if $N$ is the smallest possible number, such that the quadrature integrates exactly a particular set of functions $\{\varphi_k\}_{k=1}^M$:
\begin{equation}
\int_{\mathbb{S}^2} \varphi_k(\vec\Omega)\; d\vec\Omega=\sum_{i=1}^N w_i \varphi_k(\vec\Omega_i),\qquad 1\le k\le M\,.
\label{eq:qexact}
\end{equation}

In one dimension, the subject of quadratures is very well understood \citep{nr07,dahlquist08}. The $N$-point Gaussian quadratures are perhaps the most prominent example of the quadratures that integrate the polynomials up to the order $2N-1 $ exactly. This quadrature has also been successfully used in numerical RT. On the other hand, integration on the spherical surface, or angular integration, is still a subject of active mathematical research. The most common approach to the problem is to choose $\{\varphi_k\}_{k=1}^M$ to be a set of spherical harmonics up to a certain order $L$ \citep[e.g.][]{lebedev76,braucharta15,beentjes15}, but alternative approaches exist \citep[e.g.][]{ab09}. For more general methods of numerical integration of functions in $\mathbb{R}^d$, see \citet{xiao10}.

In polarisation transfer problems, we deal with angular dependencies of the four Stokes parameters, namely the specific intensity $I$, two components $Q$ and $U$ of the linear polarisation, and the circular polarisation $V$. These parameters can be grouped into a formal vector and indexed as $\vec (I,Q,U,V)\equiv(I_0,I_1,I_2,I_3)$. It is important to realise that the amplitude of $I_0$ is often much larger than that of the other parameters, but, in practice, we still need to be able to integrate these low-amplitude quantities to the accuracy of a few decimal places using the same quadrature.

In the case of non-axisymmetric problems, and especially in the general 3D radiative transfer, a common choice of the quadrature over the unit sphere is a product of two one-dimensional quadratures (or tensor product quadrature), namely the Gaussian quadrature in the cosine of inclinations and the equally-spaced trapezoidal quadrature in azimuths (GT quadrature in the following text). While this quadrature is easy to construct, it is not optimal due to unnecessary accumulation of points near the `poles' \citep[see Fig. 1 of][]{beentjes15}. Another popular set of quadrature rules in numerical radiative transfer, which is not of the tensor product type, is the one of \citet{carlson63}. The RT codes allowing 3D radiative transfer, such as RHSC3D \citep{uit98}, Multi3D \citep{multi3d}, and PORTA \citep{porta13}, typically implement one of these quadratures. In this paper, we do not discuss the Carlson quadratures because our calculations indicate that they fail to correctly integrate even weakly anisotropic polarised radiation fields. Instead, we compare our newly derived quadratures with the GT ones.

While in the unpolarised RT transfer theory, the important quantity is the mean radiation field
\begin{equation}
J=\int_{\mathbb{S}^2}I(\vec\Omega)\;d\vec\Omega\,,
\end{equation}
the key quantities in the polarised transfer case are the irreducible spherical tensors of the radiation field
\begin{equation}
J^K_Q=\sum_{j=0}^3\int_{\mathbb{S}^2}I_j(\vec\Omega)\mathcal{T}^K_Q(j,\vec\Omega) \;d\vec\Omega\,,
\label{eq:jkq}
\end{equation}
where $\mathcal{T}^K_Q(j,\vec\Omega)$ is the geometrical tensor defined in Table~5.6 of \citet{ll04} or Table~1 of \citet{bommier97}.

It is natural to expand the angular dependence of the Stokes parameters into the basis of spherical harmonics and to restrict the basis to a certain maximum order $L$. We then define the optimal quadrature of the order $L$ as a quadrature with a minimum number of points $N,$ which integrates exactly all of the $J^K_Q$ tensors for the Stokes parameters expansion up to the order $L$. The suitable value of $L$ depends on the problem to be solved. Our experience with the 3D non-LTE calculations in the solar atmosphere shows that $L\approx 10$ should be sufficient for such kind of applications. We leave the discussion of this topic for Sect.~\ref{sec:results} and for the following paper in this series.

There are additional constraints on the quadratures, which we require for practical reasons. Firstly, none of the rays are allowed to be parallel to any of the $XYZ$ axes of the reference frame. The formal solution of the RTE and the calculation of the ray intersections with the mesh planes is more efficient if all the rays are not parallel to any axes. Secondly, we impose the natural assumption that the quadrature is rotationally invariant with respect to $90^\circ$ rotations around the $Z$ axis and that it is mirror-symmetric with respect to the $Z=0$ plane. This way, we can restrict the search for the quadrature points to just $n=N/8$ rays in the first octant and to obtain the rest of the rays by simple rotations and/or reflection. We note that this choice of the quadrature symmetries is not the only possible one and more investigations should be done in the future regarding other constraints.


\section{Calculation of the quadratures\label{sec:calc}}

We start our derivation by expanding the Stokes parameters $\{I_j(\vec\Omega)\}_{j=0}^3$ into the basis of spherical harmonics,
\begin{equation}
I_j(\vec\Omega)=\sum_{\ell=0}^\infty\sum_{m=-\ell}^\ell (I_j)_{\ell m}Y_{\ell m}(\vec\Omega)\,,
\label{eq:ijexp}
\end{equation}
where
\begin{equation}
Y_{\ell m}(\vec\Omega)= \begin{dcases*}
        i\sqrt{2\pi}\left[ Y^m_\ell(\vec\Omega) - (-1)^m Y^{-m}_\ell(\vec\Omega) \right]  & if $m<0$\\
        2\sqrt\pi Y^0_\ell & if $m=0$\\
        \sqrt{2\pi}\left[ Y^{-m}_\ell(\vec\Omega) + (-1)^m Y^m_\ell(\vec\Omega) \right] & if $m>0$
        \end{dcases*}
\label{eq:real}
\end{equation}
are the real-valued spherical harmonics constructed from the complex-valued spherical harmonics
\begin{equation}
Y^m_\ell(\chi,\theta)=\sqrt{\frac{2\ell+1}{4\pi}\frac{(\ell-m)!}{(\ell+m)!}}P^m_\ell(\cos\theta)e^{im\chi}\,,
\end{equation}
where $P^m_\ell(\cos\theta)$ are the associated Legendre polynomials \citep{abramowitz14}. In this paper, we adopt the common definition in which the inclination angle $\theta$ is measured from the positive axis $Z$ and the azimuth $\chi$ is measured from the positive axis $X$. The spherical harmonics form an orthonormal basis for functions on $\mathbb{S}^2$. While the functions to be integrated are complex valued, we use the real spherical harmonics for expansion of the real-valued Stokes parameters. The expansion coefficients $(I_j)_{\ell m}$ can be obtained from the inverse transformation
\begin{equation}
(I_j)_{\ell m}= \int_{\mathbb{S}^2} I_j(\vec\Omega)Y_{\ell m}(\vec\Omega)\;d\vec\Omega\,.
\label{eq:ijlm}
\end{equation}

In this work, we assume that the angular dependence of $I_j(\vec\Omega)$ is such that decomposition into spherical harmonics can be truncated at a finite order $L$. With this assumption, Eq.~(\ref{eq:ijexp}) becomes
\begin{equation}
I_j(\vec\Omega)=\sum_{\ell=0}^L\sum_{m=-\ell}^\ell (I_j)_{\ell m}Y_{\ell m}(\vec\Omega)\,.
\label{eq:ijexp_trunc}
\end{equation}
This is usually a reasonable assumption in stellar photospheres and chromospheres, such as the solar ones, in which the angular dependence is not particularly peaked at some angles. In other cases of astrophysical interest, it may be desirable to derive different quadrature rules. The value of $L$ depends on the particular problem, the desired accuracy, and the constraints on the CPU time. In general, the larger $L$ is, the higher the accuracy is of the numerical solution.

Using the expansion of $I_j(\vec\Omega)$ from Eq.~(\ref{eq:ijexp_trunc}) in Eq.~(\ref{eq:jkq}), we obtain the expression for the radiation field tensor in terms of the Stokes-parameter coefficients
\begin{equation}
J^K_Q=\sum_{j=0}^3\sum_{\ell=0}^L\sum_{m=-\ell}^\ell (I_j)_{\ell m} y(j,\ell, m, K ,Q)\,,
\label{eq:jkq1}
\end{equation}
where the complex-valued factors
\begin{equation}
y(j,\ell, m, K ,Q)=\int_{\mathbb{S}^2} Y_{\ell m}(\vec\Omega)\mathcal{T}^K_Q(i,\vec\Omega)\;d\vec\Omega
\end{equation}
can be analytically calculated for any combination of the indices. From the definition of the quadrature in Eq.~(\ref{eq:qdef}) and by applying the expansion of the Stokes parameters, we can express $J^K_Q$ as
\begin{equation}
J^K_Q=\sum_{j'=0}^3\sum_{i=1}^N w_i I_{j'}(\vec\Omega_i)\mathcal{T}^K_Q(j',\vec\Omega_i)
=\sum_{j'=0}^3\sum_{\ell'=0}^L\sum_{m'=-\ell'}^{\ell'} (I_{j'})_{\ell'm'}\sum_{i=1}^N w_i Y_{\ell'm'}(\vec\Omega_i)\mathcal{T}^K_Q(j',\vec\Omega_i)\,.
\label{eq:jkq2}
\end{equation}
Since the right-hand sides of Eqs.~(\ref{eq:jkq1}) and (\ref{eq:jkq2}) must be equal for any realisation of the radiation field, that is, for any values of $(I_j)_{\ell m}$ and $(I_{j'})_{\ell' m'}$, it follows that the coefficients multiplying these radiation field components must be the same in both equations, that is,
\begin{equation}
y(j,\ell,m,K,Q)=\sum_{i=1}^N w_i Y_{\ell m}(\vec\Omega_i)\mathcal{T}^K_Q(j,\vec\Omega_i)
\label{eq:qeqs}
\end{equation}
for any set of indices. Any quadrature, which satisfies the set of equations (\ref{eq:qeqs}), integrates exactly the $J^K_Q$ tensors for any realisation of the radiation field. The optimal quadratures up to the order $L$ are the ones with the minimum value of $N$. The functions $Y_{\ell m}(\vec\Omega)\mathcal{T}^K_Q(j,\vec\Omega)$ play the role of the functions $\varphi_k$ in Eq.~(\ref{eq:qexact}).

The set of equations (\ref{eq:qeqs}) for the unknown values of $w_i$ and directions $\vec\Omega_i$ is non-linear and can only be solved by numerical methods. Moreover, given $L$, the value of $N(L)$ is also unknown. In practice, knowing the value of $n(L-1)=N(L-1)/8$, we start searching for $n(L)=n(L-1)$ and, in case no solution is found, we proceed by increasing $n(L)$ by 1 until the quadrature is found. Apart from $n$ unknown weights $w_i$, the unknown directions are represented by the inclination $\theta_i\in(0^\circ,90^\circ)$ and azimuths $\chi_i\in(0^\circ,90^\circ)$ in the first octant defined by the positive axes $X$, $Y$, and $Z$. Given the necessarily numerical solution of the problem, the exactness of the quadrature is limited by the double-precision computer arithmetics. Our condition for the quadrature to be considered exact is
\begin{equation}
\underset{j,\ell,m,K,Q}{\max}
\left\| y(j,\ell,m,K,Q)-\sum_{i=1}^N w_i Y_{\ell m}(\vec\Omega_i) \mathcal{T}^K_Q(j,\vec\Omega_i) \right\|< 10^{-15}\,,
\end{equation}
that is, the maximum error of integration of any of the $Y_{\ell m}(\vec\Omega)\mathcal{T}^K_Q(j,\vec\Omega)$ functions is smaller than $10^{-15}$. We solved Eqs.~(\ref{eq:qeqs}) as a least-squares minimisation problem using the trust region reflective algorithm \citep[TRF,][]{branch99}. The initial guess of the quadrature is chosen randomly, but since the problem is non-convex, that is, it suffers from the existence of local minima, it is useful to start with a guess that is as close to the global minimum as possible. Therefore, the initial guess used as a starting point for the TRF method is chosen as the most precise one out of the number of randomly generated quadratures. It can be easily shown that thanks to the symmetry of the quadrature with respect to rotation by integer multiples of $90^\circ$ around the $Z$ axis, a significant number of Eqs.~(\ref{eq:qeqs}) are identically satisfied, which simplifies the numerical solution.

In the case of unpolarised radiative transfer, the problem is significantly simplified. The only non-zero Stokes parameter is the specific intensity $I_0$ and the only relevant geometrical tensor is $\mathcal{T}^0_0(0,\vec\Omega)=1$. Therefore, the Eqs.~(\ref{eq:qeqs}) take the following form:
\begin{equation}
y(0,\ell,m,0,0)=\sum_{i=1}^N w_i Y_{\ell m}(\vec\Omega_i)\,.
\end{equation}
The optimal quadrature for unpolarised RT is then simply the one that is optimal for the integration of spherical harmonics.


\section{Results\label{sec:results}}

\begin{figure}
\begin{center}
\includegraphics[width=0.7\columnwidth]{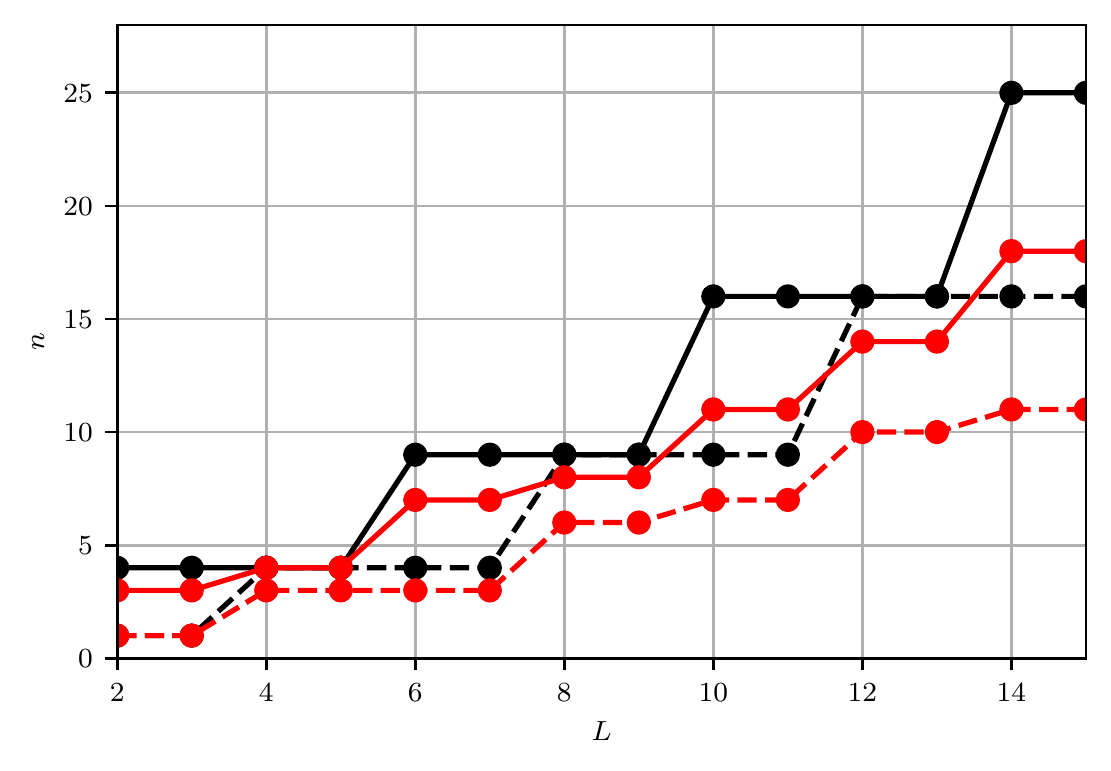}
\end{center}
\caption{
Comparison of the number of quadrature points per octant, $n(L)=N(L)/8$, of the GT quadratures (black lines) and the new quadratures (red lines) as a function of the order $L$. The solid lines correspond to the full-Stokes case and the dashed lines show the unpolarised (Stokes $I$) quadratures, respectively.
}
\label{fig:gtcomp}
\end{figure}

\begin{figure}
\begin{center}
\includegraphics[width=0.7\columnwidth]{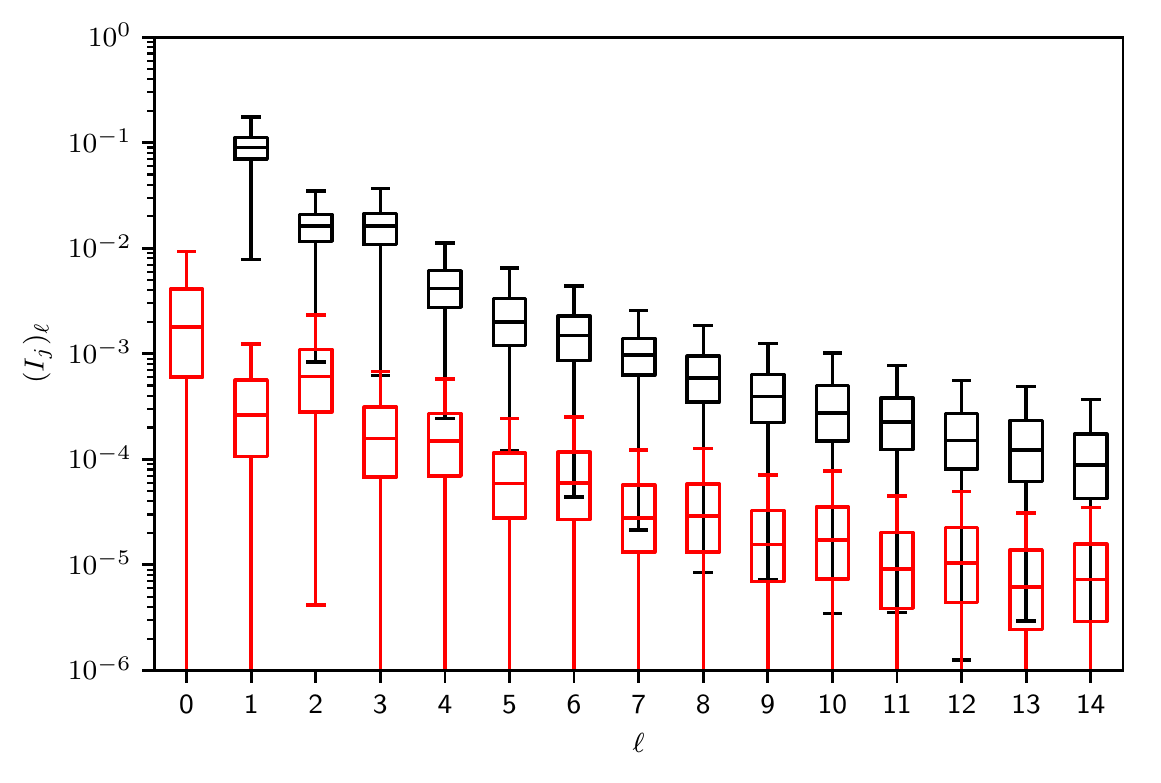}
\end{center}
\caption{
Statistical distribution of the amplitudes $(I_j)_\ell$ defined in Eq.~(\ref{eq:ijl}) of the spherical harmonic components of Stokes $I$ (black) and Stokes $Q$ (red) at the unit optical depth for disc-centre line of sight of the Ca\,{\sc i}\,4227\,\AA\ line in a 3D snapshot model of the solar atmosphere \citep{carlsson16} calculated using a very fine GT quadrature ($20 \times 20$ points per octant).
}
\label{fig:ls}
\end{figure}

\begin{figure}
  \begin{center}
    \includegraphics[width=0.45\columnwidth]{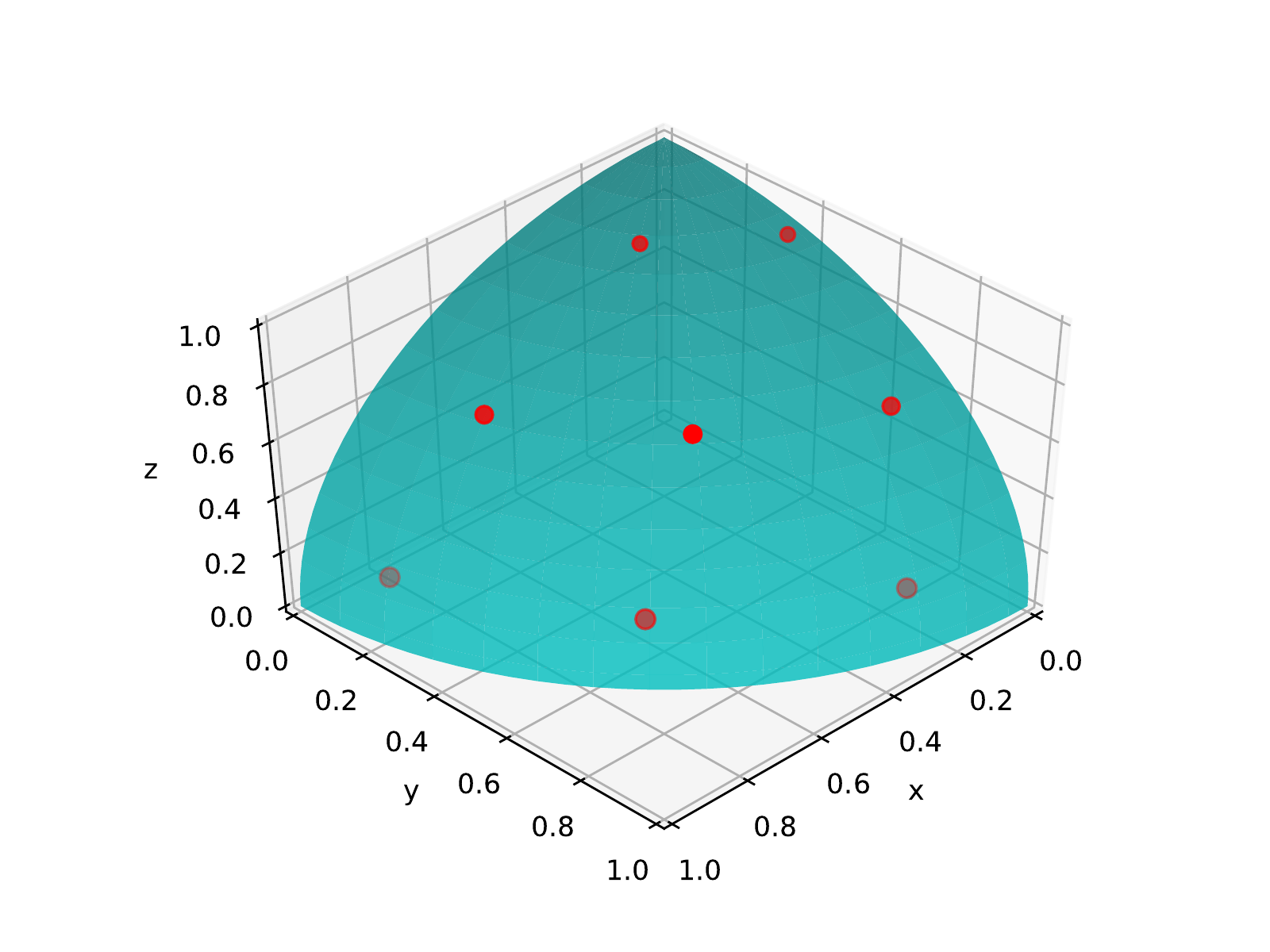}
    \includegraphics[width=0.45\columnwidth]{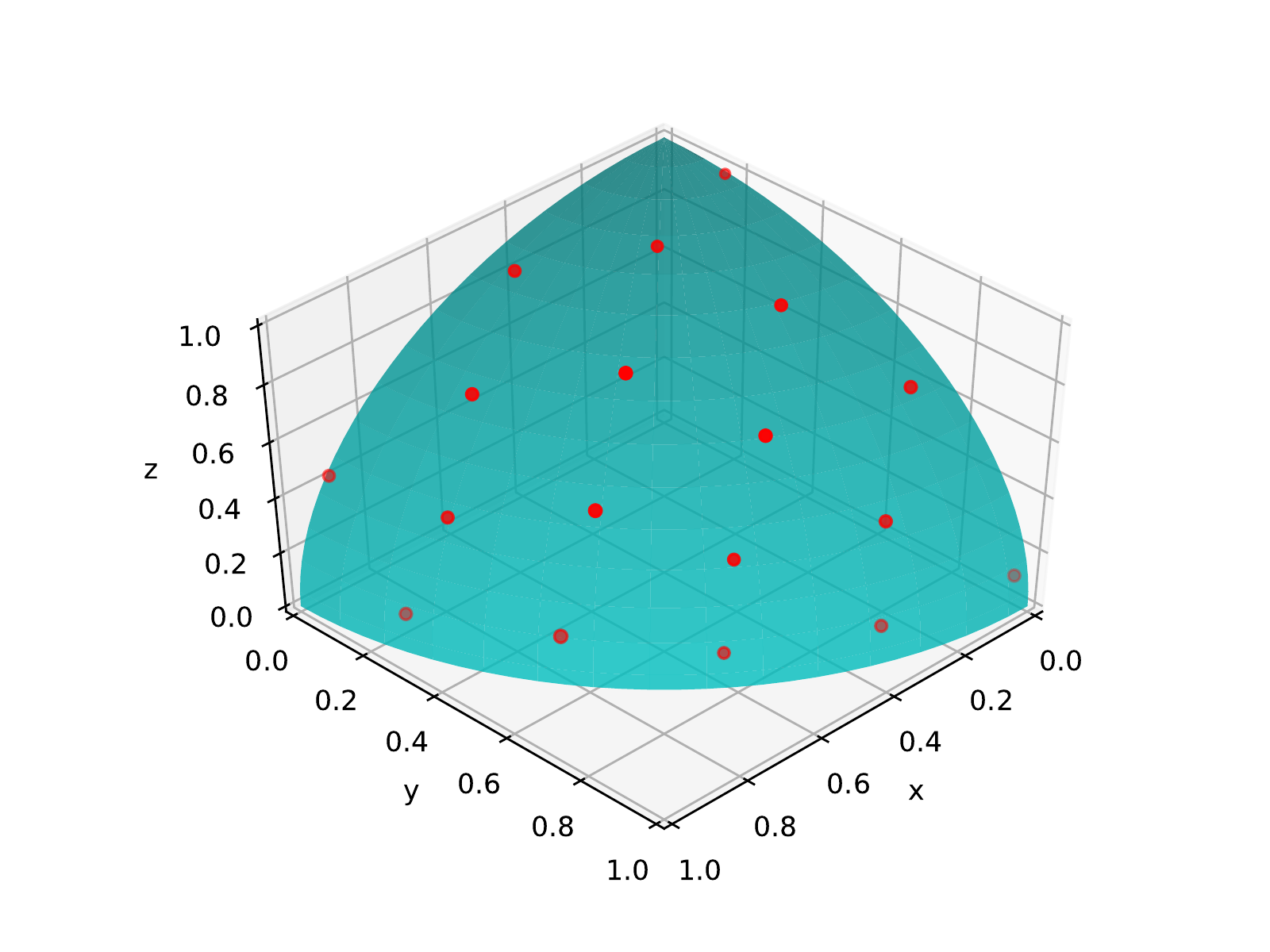}
    \caption{Distribution of the quadrature points over the unit sphere in the first octant for the case of the $L=9$ (left) and $L=15$ (right) quadratures. The red areas of the points are proportional to their weights.}
    \label{fig:quads}
  \end{center}
\end{figure}

We derived the quadrature rules with an increasing accuracy up to $L=15$. The number of these near-optimal quadrature points per octant as a function of $L$ is shown in Fig.~\ref{fig:gtcomp}. For comparison purposes, we also show the number of points per octant in the GT quadrature needed for exact integration up to the same order.

It follows from the figure that with the exception of very small orders, the new quadratures perform better than the GT ones. For some orders, the number of rays is about $30\,\%$ smaller at the same level of accuracy. A similar behaviour is found for the unpolarised case, which is shown with the dashed lines.

In order to provide some insight as to how to choose $L$ in practical calculations, we define the quantity
\begin{equation}
(I_j)_\ell = \frac{1}{2\ell+1} \sqrt{ \frac{\sum_{m=-\ell}^{\ell} (I_j)_{\ell m}^2}{(I_0)_{00}^{2}} } \,,
\label{eq:ijl}
\end{equation}
which characterises, at a given point in the 3D model, the expansion coefficients of the given order $\ell$ for each Stokes parameter. In Fig.~\ref{fig:ls} we show box plots of $(I_j)_\ell$ in a particular 3D model of the solar atmosphere. The coefficients decrease approximately exponentially with $\ell$; in other words, the role of the coefficients rapidly decreases with their order. An important fact to realise is that the anisotropy of the specific intensity often dominates some of the $J^K_Q$ components, in part, because the intensity is usually significantly higher than the other Stokes parameters. In order to accurately account for the contribution of all of the Stokes parameters, it is important to choose $L$ such that the amplitudes $(I_0)_L$ of the intensity are smaller than at least the first amplitudes $(I_j)_\ell$ with $j>0$ of the other Stokes parameters. In the case of the Ca\,{\sc ii}\,4227\,\AA\ line, this means $L \gtrapprox 6$.

In Figure \ref{fig:quads} we show how the rays are distributed in the $L=9$ and $L=15$ quadratures. The rays are more evenly distributed over the unit sphere in contrast to the GT quadrature, hence we avoided oversampling in some regions.

In Appendix~\ref{sec:app}, some of the new quadratures with different levels of accuracy are tabulated. All the new quadratures indicated in Fig.~\ref{fig:gtcomp} can be found in the on-line material.


\section{Discussion and conclusions\label{sec:concl}}

We derived a set of new quadratures for the angular integration of polarised radiation fields. As a criterion for optimality, we used the condition of exact integration of the $J^K_Q$ radiation field tensors up to a given order of expansion of the Stokes parameters in the basis of spherical harmonics. For the maximum orders up to $L=15$ that we studied, our quadratures can save up to 30\,\% of ray directions, hence approximately 30\,\% of computing time compared to the GT quadratures of the same accuracy.

Our derivation of near optimal angular quadratures was based on the $J^K_Q$ tensors that appear in the complete frequency redistribution (CRD) theory of spectral line formation \citep[see][]{ll04}. In the general case of partial redistribution (PRD), other quadratures may prove to be more suitable. We also note that in some situations, such as the case of plasma structures embedded in outer stellar envelopes, it may be advantageous to consider different quadratures to properly quantify the impact of the highly anisotropic illumination of the structure from the stellar surface. This type of investigation is out of the scope of the present paper.

Due to the non-linearity and non-convexity of the problem, there is not a unique solution to the problem. Moreover, we cannot prove analytically that the resulting quadratures are really optimal in terms of the minimal number of quadrature points, hence we can only claim that our empirical approach provides nearly optimal quadratures. Our numerical experiments indicate that quadratures with different symmetry properties with respect to rotations and reflections of the coordinate system may be of future interest in some applications. In the following paper, we will demonstrate a practical comparison of the new quadratures with the GT quadratures in realistic 3D non-LTE radiative transfer models of the solar atmosphere.


\begin{acknowledgements}
J.\v{S}. acknowledges the financial support of the grant \mbox{19-20632S} of the Czech Grant Foundation (GA\v{C}R) and from project RVO:67985815 of the Astronomical Institute of the Czech Academy of Sciences. J.T.B. acknowledges the funding received from the European Research Council (ERC) under the European Union’s Horizon 2020 research and innovation programme (ERC Advanced Grant agreement \mbox{No.~742265}).
\end{acknowledgements}


\appendix

\section{Tables of quadratures\label{sec:app}}

As an example of the quadratures derived in this paper, here we show three quadratures of increasing accuracy for the full set of Stokes parameters and the $J^K_Q$ tensors (see Table\,\ref{table:iquv}). We also show one quadrature for just the specific intensity (Table\,\ref{table:i}).\footnote{We note that according to the results of Sect.~\ref{sec:calc}, any quadrature that is capable of exact integration of the spherical harmonics up to the order $L$ can be used for the angular integration of the intensity.} The tables of the other quadratures shown in Fig.~\ref{fig:gtcomp} can be found in the on-line material of this paper.

The $n=N/8$ quadrature points in the tables correspond to the points in the first octant. The points in the remaining $Z>0$ octants were generated by rotating the first-octant points by 90, 180, and 270 degrees, while preserving their weights. The points in the $Z<0$ octants were generated by reflection of all the $Z>0$ points with respect to the $Z=0$ plane and by preserving their weights.

\begin{table*}
\caption{Full-Stokes quadratures.}
\label{table:iquv}
\centering
\begin{tabular}{llll}
\hline\hline
$w_i$ & $\theta_i\;[^\circ]$ & $\chi_i\;[^\circ]$\\
\hline
\noalign{\smallskip}
\multicolumn{3}{c}{$L=5$, $n=4$} \\
\noalign{\smallskip}
\hline
0.03008835383493305  &  76.52405222057945  &  60.76839852319566  \\
0.042736793710787514  &  69.12677915845973  &  15.768398523195653  \\
0.02500191721861294  &  24.668834456356183  &  15.768398523195646  \\
0.027172935235666495  &  44.21094605227794  &  60.76839852319564  \\
\hline
\noalign{\smallskip}
\multicolumn{3}{c}{$L=11$, $n=11$} \\
\noalign{\smallskip}
\hline
0.013928998075380375  &  76.93315550646489  &  80.7688199429568  \\
0.011639454305572013  &  50.1602530026253  &  35.76881994295678  \\
0.011029475400118194  &  57.530476539597316  &  58.26881994295681  \\
0.01117777914840798  &  79.76232528690839  &  58.26881994295679  \\
0.011639454305572  &  50.1602530026253  &  80.76881994295682  \\
0.011499122313789287  &  59.67896625851446  &  13.268819942956803  \\
0.012557626403854785  &  28.786346749828134  &  58.26881994295679  \\
0.011968777966644249  &  30.89607258580027  &  13.268819942956805  \\
0.005651235410063879  &  8.118127482350042  &  13.268819942956803  \\
0.013928998075380379  &  76.93315550646489  &  35.768819942956796  \\
0.009979078595216876  &  81.29512958261202  &  13.268819942956807  \\
\hline
\noalign{\smallskip}
\multicolumn{3}{c}{$L=15$, $n=18$} \\
\noalign{\smallskip}
\hline
0.00591402443445905  &  20.6600328200269  &  42.946184169224324  \\
0.008596636090469512  &  80.01111744010231  &  33.688427806802814  \\
0.007113673482367161  &  64.92199506534526  &  72.01046766489479  \\
0.006827258788941489  &  64.14751339655844  &  18.497227186025057  \\
0.006754747098551407  &  64.73031971805044  &  0.91953444887696  \\
0.007479599563267462  &  39.67406768031979  &  38.668043251384816  \\
0.007204332706475753  &  47.769294855310704  &  83.11145268513937  \\
0.004857493646975331  &  10.141119789523483  &  85.35219402354653  \\
0.007336667967350494  &  81.51014220303644  &  87.34267597224692  \\
0.0070868921059280676  &  81.78582676152016  &  69.36594366874719  \\
0.006578810918075098  &  66.80381195280124  &  53.058440530090046  \\
0.006617138696419558  &  28.21168023676325  &  9.168537444166633  \\
0.007781933481283711  &  59.472237532728144  &  36.55217214395332  \\
0.006596638597620959  &  31.588144273444552  &  69.31235802205393  \\
0.007274106171534564  &  49.28592372130238  &  59.21387340583087  \\
0.007120519412160358  &  46.20368511852779  &  15.621163533242104  \\
0.007306722626653114  &  81.72467073554755  &  15.505757262020685  \\
0.006552804211466908  &  82.37095667966659  &  51.47786543375931  \\
\hline\\
\end{tabular}
\end{table*}

\begin{table*}
\caption{Stokes-$I$ quadrature for $L=9$, $n=6$.}
\label{table:i}
\centering
\begin{tabular}{llll}
\hline\hline
$w_i$ & $\theta_i\;[^\circ]$ & $\chi_i\;[^\circ]$\\
\hline\\
0.022713607613686692  &  42.88891950700113  &  25.472196549014534  \\
0.022787199039712287  &  73.84377826798516  &  9.414172189274723  \\
0.022110192033907447  &  74.48396537369167  &  40.11547359112626  \\
0.022654670920532104  &  47.784764086172665  &  69.27004143167122  \\
0.018155933946312444  &  77.01768621593254  &  69.815112660868  \\
0.01657839644584903  &  18.072787166919895  &  83.12883371332764  \\
\hline\\
\end{tabular}
\end{table*}


\end{document}